\documentclass{ws-procs975x65}

\begin{document}

\title{REPORT ON SESSION QG4 OF THE 13TH MARCEL GROSSMANN MEETING}

\author{JORGE PULLIN}

\address{Department of Physics and Astronomy, Louisiana State
  University, Baton Rouge, LA 70803\\
pullin@lsu.edu}

\author{PARAMPREET SINGH}

\address{Department of Physics and Astronomy, Louisiana State
  University, Baton Rouge, LA 70803\\
psingh@lsu.edu}

\begin{abstract}
We summarize the talks presented at the QG4 session (loop quantum
gravity: cosmology and black holes) of the 13th Marcel Grossmann
Meeting held in Stockholm, Sweden.
\end{abstract}



\bodymatter

\vskip1cm

The session was devoted to results on black holes and cosmology in the
context of loop quantum gravity. It was a vibrant session where
several new advances were highlighted. Among them, a much clearer
picture emerges of the computation of the entropy in loop quantum
gravity and several new insights into singularity resolution and how
to treat time dependent quantizations arise in the context of
cosmologies. Here is a summary of the talks presented:

Carlo Rovelli reviewed and provided context for certain important
developments in the calculation of the entropy of horizons in loop
quantum gravity. He started with the Unruh temperature (Ref.~\refcite{unruh}),
the results of Kubo, Martin and Schwinger (Ref.~\refcite{kms}) , discussed
results of Tolman concerning gravitational thermodynamics and then
proceeded to discuss the connection with the talks by Pranzetti, Perez
and Bianchi in this session.

Alejandro Perez presented results obtained recently with Amit Ghosh
and Ernesto Frodden and discussed in
Ref.~\refcite{pegofr}. Essentially they show that stationary black
holes satisfy a very simple local form of the first law using a
preferred family of local observers near the horizon and a suitable
definition of energy for them. The same construction can be applied in
the context of isolated horizons. When applied in the framework of
loop quantum gravity, and treating the number of punctures in the
horizon as a non-trivial observable, this leads to a grand canonical
calculation that agrees with Hawking's semiclassical analysis for all
values of the Immirzi parameter. It appears that matter inclusion is
inevitable in the local calculation. And this could explain the flux
of the entropy from its ultraviolet value (different from Hawking's
calculation) to its infrared one (which coincides with Hawking's).

Eugenio Bianchi discussed several considerations concerning black hole
entropy. Among them the use of the spin foam action to provide a
classical theory with boundaries and the states that appear in the
horizon dynamics in terms of spin foams (discussed in
Ref.~\refcite{bianchi1}) and the quantum Rindler horizon, how to
assign an energy to the quantum horizon that reproduces the
Frodden--Ghosh--Perez result, how to assign a temperature to the
quantum horizon that leads to the usual result and how to use the
Clausius relation to obtain the entropy, again leading to the usual
result without fixing the Immirzi parameter.

Fernando Barbero discussed work in collaboration with Eduardo
Villase\~nor, presented in Ref.~\refcite{barberovilla}, concentrating
on the subdominant corrections to black hole entropy. In particular it
is observed that the corrections predicted by different models are
different, their sign (which is relevant for black hole stability) is
ensemble-dependent.

Jacobo D\'{\i}az-Polo summarized work with Aurelien Barrau, Thomas
Cailleteau, Xiangyu Cao and Julien Grain presented in
Ref.~\refcite{barrau} concerning the imprint that is left in the
Hawking radiation by the structure of the black hole area spectrum
emerging in loop quantum gravity and how this could be used to
differentiate it from other theories through the observation of black
hole radiation.

Rodolfo Gambini discussed work with N\'estor \'Alvarez, Saeed Rastgoo
and Jorge Pullin concerning the quantization of spherically symmetric
space-times coupled to a scalar field in loop quantum gravity. This
requires a strategy to deal with the non-trivial algebra of
constraints that emerges in this example. Strategies discussed
included the use of uniform discretizations, discussed in
Ref.~\refcite{saeed} and a recently introduced family of gauge fixings
that lead to a local true Hamiltonian presented in
Ref.~\refcite{true}.

Daniele Pranzetti analyzed what happens when one matches the dynamical
horizons framework with the local thermodynamical approach of Frodden,
Ghosh and Perez. It allows to study the radiation process generated by
the loop quantum gravity dynamics near the horizon, providing a
quantum gravity description of the black hole evaporation. For large
black holes it leads to a spectrum with a discrete structure that
could be potentially observable. The results are discussed in
Ref.~\refcite{pranzetti}

Ernesto Frodden discussed work with Alejandro Perez, Daniele Pranzetti
and Christian R\"oken concerning the quantum origin of the entropy of
rotating black holes. The approach is to consider a classical theory
with an isolated horizon and identify the symplectic structure and in
the quantization use a Chern--Simons topological theory with defects
at the boundary.

Ivan Agull\'o presented a quantum gravity extension of the
inflationary scenario obtained in collaboration with Abhay Ashtekar
and William Nelson and discussed in Ref.~\refcite{agullo}. The key
idea is to analyze quantum fluctuations over quantum spacetime whose
homogeneous part is quantized using loop quantum cosmology. This
approach overcomes two major difficulties of the conventional
inflationary paradigm: the singularity and the trans-Planckian
problems. It is found that for a large class of initial conditions
cosmological perturbations yield results consistent with the WMAP
data. Interestingly, there exists a narrow range of parameters for
which distinct signatures from standard inflation are also possible.
These results, thus provide an opportunity to test predictions of loop
quantum gravity in future astronomical observations.

Edward Wilson-Ewing discussed about lattice loop quantum cosmology --
an approach to go beyond the homogeneity assumption in loop quantum
cosmology, presented in Ref.~\refcite{edward}. In this approach,
spacetime is divided in to homogeneous and isotropic cells which are
coupled with each other. The scalar constraint in this approach is
composed of an ultralocal homogeneous term and an interaction term,
which along with the diffeomorphism constraint, turns out to be
preserved by the dynamics for small perturbations. Using the effective
dynamics resulting from quantum constraints, loop quantum effects on
linear perturbations can be studied.

Tomasz Pawlowski discussed a way the geometric degrees of freedom can
be used for deparameterization in loop quantum cosmology. Pawlowski
analyzed the case of a massive scalar field using volume and its
momentum deparameterization, and reported the properties of quantum
evolution operator and its eigenfunctions. The approach aims to
overcome some difficulties which may arise in treating matter degrees
of freedom as internal time.  Preliminary results seem to suggest a
non-trivial contribution to the energy density of the matter which may
pose difficulties for renormalization for infinite modes.

Thomas Cailleteau discussed results on some observational consequences
of loop quantum cosmology obtained in collaboration with Aurelien
Barrau, A. Demion, Julien Grain, Jakub Mielczarek and Francesca
Vidotto, and presented in Ref.~\refcite{obs}. These results are based
on the approach of considering an effective Hamiltonian which captures
inhomogeneous degrees of freedom in loop quantum cosmology. The
effective Hamiltonian incorporates the key effects which are expected
to originate from the underlying quantum geometry. Effective
constraints in this approach lead to anomalies, which require
introduction of counter terms for their cancellation. With the inclusion
of these counter terms, it is possible to obtain more consistent
constraints on phenomenological parameters and a better understanding
of some conceptual issues using the analysis of cosmological
perturbations. The results are consistent with those of Edwin
Wilson-Ewing.

Mikel Fern\'andez-M\'endez presented a hybrid quantization of an
inhomogeneous inflationary model in loop quantum cosmology done in
collaboration with Guillermo Mena-Marug\'an and Javier Olmedo and
presented in Ref.~\refcite{hybrid1}. In the hybrid quantization
approach, the key idea is to quantize homogeneous degrees of freedom
using loop quantum gravity techniques, and treat inhomogeneities as
Fock quantized perturbations over the homogeneous background.  By
fixing the gauge for local degrees of freedom at the classical level,
properties of the quantum Hamiltonian constraint for the scalar
perturbations were reported.

Daniel Martin de Blas discussed an approximation scheme developed in
collaboration with Mercedes Mart\'{\i}n-Benito and Guillermo
Mena-Marug\'an to study inhomogeneities in loop quantum cosmology for
the hybrid quantization of Gowdy $T^3$ model and Bianchi-I model which
are locally rotationally symmetric.  The quantization of both systems
is performed including a massless scalar field. For the Gowdy model,
the quantum Hamiltonian constraint can be written as a sum of the
constraint for the homogeneous and isotropic spacetime and anisotropic
and interaction terms arising due to inhomogeneity. The involved
approximation, that concerns with considering eigenstates of the FRW
operator, allows one to obtain approximated solutions. For the Bianchi
I model, a solvable quantum Hamiltonian constraint is obtained, being
possible to construct semi-classical solutions and study their
evolution. 

Francesca Vidotto reviewed basics of spinfoams for cosmologists, and
some of the key results which have been obtained so far in spinfoam
cosmology and discussed in Ref.~\refcite{vidotto}. The idea is to
choose a suitable graph capturing inhomogeneous degrees of freedom and
use coherent state techniques to compute transition amplitudes between
two states of a universe. Under certain approximations, using this
transition amplitude, an effective Hamiltonian capturing the quantum
gravity corrections can be obtained.

Andrea Dapor gave two talks. First of these talks was on a joint work
with Michal Artymowski and Tomasz Pawlowski on a non-minimally coupled
inflationary model in loop quantum cosmology and presented in
Ref.~\refcite{dapor1}. The authors analyzed dynamics in the Planck
regime and the effects of non-minimal coupling using an effective
Hamiltonian.  In the second talk, Andrea Dapor presented results on
quantum field theory on Bianchi-I spacetime in loop quantum cosmology
obtained in collaboration with Jerzy Lewandowski and Yaser Tavakoli
and presented in Ref.~\refcite{dapor2}. These results extend the
previous work in loop quantum cosmology on quantum field theory in
quantum Friedmann-Robertson-Walker spacetime to an anisotropic
setting. By obtaining an effective spacetime geometry, Dapor and
colleagues studied Lorentz invariance violation. They found that it
may be possible for Lorentz violation to occur when the backreaction
of quantum fields on quantum spacetime is included.

Jos\'e Velhinho discussed uniqueness of Fock quantization of scalar
fields with time dependent mass presented in
Ref.~\refcite{velhinho}. These results obtained in collaboration with
Jer\'onimo Cortez, Daniel Mart\'{\i}n-de Blas, Laura Gomar, Guillermo
Mena Marug\'an, Mikel Fern\'andez-M\'endez and Javier Olmedo aim to
shed insights on the ambiguities in the choice of Fock representation
for the canonical commutation relations in quantum field theory in
curved spacetime. Velhinho discussed that using unitary dynamics
criteria and exploiting underlying spatial symmetries, a unique
unitary equivalence class of Fock quantizations can be chosen.

Francesco Cianfrani discussed an approach to consider inhomogeneous
cosmological spacetimes in loop quantum gravity developed in
collaboration with Emanuele Alesci and presented in
Ref.~\refcite{alesci}. In this approach the goal is to obtain a
quantum cosmological model which can be obtained from loop quantum
gravity with a proper reduction. To achieve this, an inhomogeneous
Bianchi line element is considered and its quantization is proposed by
a projection of the kinematical Hilbert space of loop quantum gravity
to an appropriate subspace.

\section*{Acknowledgments}

This work was supported in part by grant NSF-PHY-0968871,
NSF-PHY-1068743, funds of the Hearne Institute for Theoretical
Physics and CCT-LSU. This publication was made possible through the
support of a grant from the John Templeton Foundation. The opinions
expressed in this publication are those of the author(s) and do not
necessarily reflect the views of the John Templeton Foundation.




\end{document}